\documentclass[aps,groupeaddress,floatfix,reprint]{revtex4-2}

\usepackage[utf8]{inputenc}
\usepackage{color}
\usepackage{bm}
\usepackage{amsmath,amssymb,amsthm}
\usepackage[colorlinks=true,urlcolor=blue]{hyperref}
\usepackage{graphicx}
\usepackage{caption,subcaption}
\usepackage{siunitx}

\begin{document}

\title{Supersymmetry-enhanced Stark-Chirped Rapid-Adiabatic-Passage in multimode optical waveguides}
\author{David Viedma}
\email{david.viedma@uab.cat}
\affiliation{Departament de F\'{i}sica, Universitat Aut\`{o}noma de Barcelona, E-08193 Bellaterra, Spain}
\author{Ver\`{o}nica Ahufinger}
\affiliation{Departament de F\'{i}sica, Universitat Aut\`{o}noma de Barcelona, E-08193 Bellaterra, Spain}
\author{Jordi Mompart}
\affiliation{Departament de F\'{i}sica, Universitat Aut\`{o}noma de Barcelona, E-08193 Bellaterra, Spain}
\date{September 2, 2021}

\begin{abstract}
We propose a method to efficiently pump an excited mode of a multimode optical waveguide starting from a fundamental-mode input by combining Stark-Chirped Rapid Adiabatic Passage (SCRAP) and Supersymmetry (SUSY) transformations. In a two-waveguide set, we implement SCRAP by modulating the core refractive index of one waveguide, which is evanescently coupled to its SUSY partner. SCRAP provides an efficient transfer of light intensity between the modes of different waveguides, while SUSY allows to control which modes are supported. Using both techniques allows to achieve fidelities above 99\% for the pumping of the excited mode of a two-mode waveguide. Additionally, we show that SCRAP can be exploited to spatially separate superpositions of fundamental and excited modes, and how SUSY can also improve the results for this application.
\end{abstract}

\maketitle

\section{Introduction}

Multimode optical waveguides and fibers \cite{Olshansky1979} provide a solution to the increasing demand of transmission capacity for optical devices \cite{Agrell2016}. Compared with single-mode waveguides, they offer an effective additional dimensionality to the system by allowing to use each guided mode as an independent information channel. This is known as Mode-Division Multiplexing (MDM) \cite{Richardson2013,Li2018,Xu2020,Su2021}, one of the different Multiplexing techniques that have been developed to increase the transmission possibilities of these optical devices \cite{Winzer2014}. The advantages that MDM provides, however, are limited by the challenging need to precisely excite the desired guided modes by matching the input field to the modes' spatial profiles. Although several techniques for this purpose exist, they either rely on carefully shaping the input pulse through e.g. spatial light modulators or phase plates \cite{Fazea2018}, or on a specific design for the propagating medium with e.g. optical fibers featuring gratings or multiple cores \cite{Richardson2016, Fazea2018}. In that sense, the possibility of precisely pumping excited modes of general multimode waveguides without the need of complicated input fields is of high interest for current and future applications.

A method that can be used to transfer light into the excited modes of a multimode waveguide is Stark-Chirped Rapid-Adiabatic-Passage (SCRAP). The method was first introduced in Ref.~\cite{Yatsenko1999} for the transfer of population to a metastable atomic state using a two-pulse scheme, and has since then been extensively studied in several cases \cite{Rickes2000,Myslivets2002,Rickes2003} and generalized to three-level systems \cite{Rangelov2005,Chang2007,Oberst2007,Shirkhanghah2020}. A strong off-resonant Stark pulse modifies the energies of a two-level system, producing two energy crossings. A short but intense pump pulse then adiabatically drives the population from the initial to the target state during only one of these crossings, while remaining negligible during the other. This scheme can be implemented in optical waveguides setups to achieve faithful transfer of light between guided modes. In this case, the role of the energy eigenvalues is played by the propagation constants of the modes, whose variation is achieved through a modulation of the refractive index of the waveguide core. The transfer is then controlled by the coupling strength between the waveguide modes, which can be tuned by adjusting the relative distance between waveguides.

In the context of optics, Supersymmetry (SUSY) has been successfully employed for instance to control the modal content of different structures \cite{Miri2013,Miri2013A,heinrich2014,principe2015,Queralto2017,Macho2018,Queralto2018,Walasik2018,Contreras2019}, their scattering properties \cite{Longhi2010,heinrich2014OL,Miri2014,Longhi2015,Garcia2020} and their topology \cite{Queralto2020}. For multimode optical waveguides, one can obtain a superpartner refractive index profile whose guided modes share propagation constants with the original waveguide, but where the fundamental mode is removed from the spectrum. In conjunction with SCRAP, this modal control allows to achieve transfer to only specific modes in a multimode waveguide, and thus be able to pump them efficiently.

Our objective is to show how the combination of SCRAP and SUSY can be exploited to faithfully pump excited modes of multimode waveguides and to spatially separate superpositions of different modes, both challenges of high importance in the context of MDM. The work is organized as follows: In section~\ref{s-theory}, we present the theoretical basis behind SCRAP and SUSY. After that, in section~\ref{s-implementation} we describe their implementation using multimode optical waveguides. We display our results in section~\ref{s-results}, where we report the fidelities of the above-mentioned applications, and finally we lay out our conclusions in section~\ref{s-conclusions}.

\section{Theory} \label{s-theory}

\subsection{Stark-Chirped Rapid-Adiabatic-Passage}

\begin{figure}[b]
    \centering
    \begin{subfigure}[t]{0.85\columnwidth}
    \includegraphics[width=\textwidth]{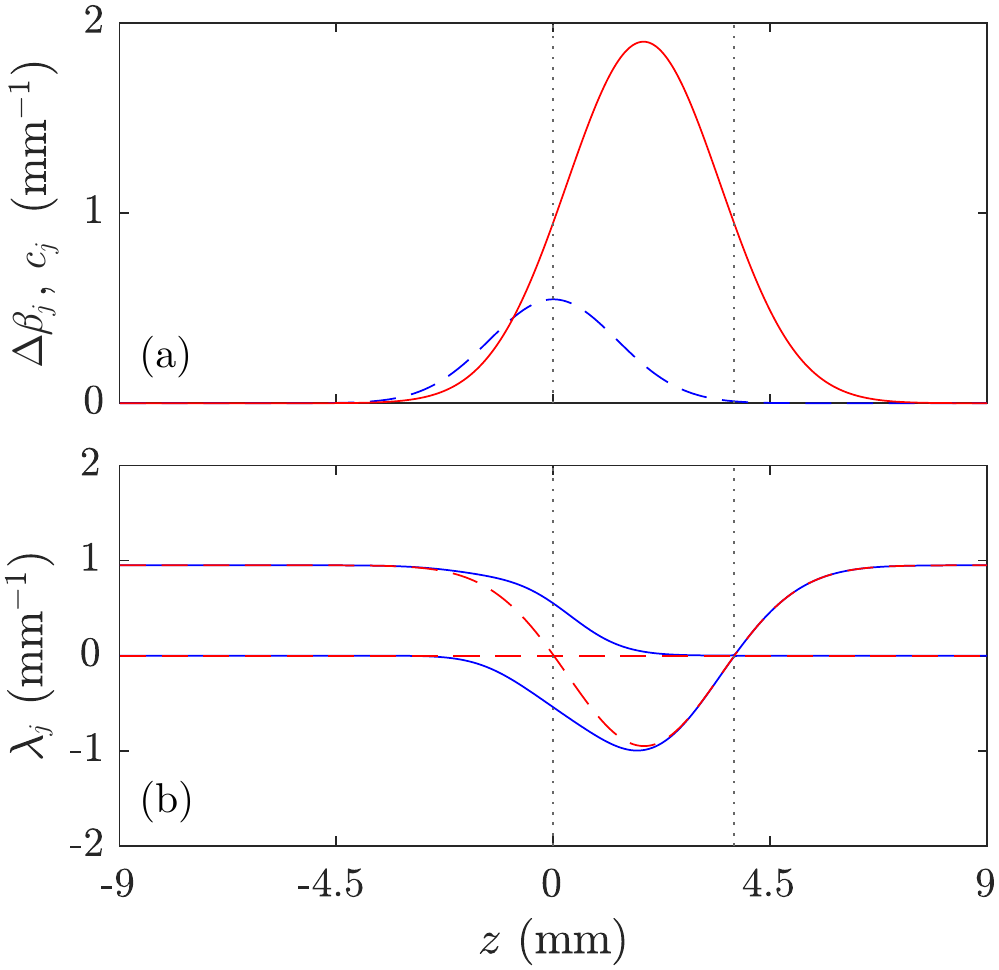}
    \phantomcaption
    \label{fig:SCRAP-pulses}
    \end{subfigure}
    \begin{subfigure}[t]{0\columnwidth}
    \includegraphics[width=0\textwidth]{example-image-b}
    \phantomcaption
    \label{fig:SCRAP-energy}   
    \end{subfigure}
    \caption{\textbf{(a)} Strength of the spatial modulation of the detuning (red solid line) and of the coupling (blue dashed line) between modes $j$ in each waveguide. \textbf{(b)} Variation of the eigenvalues (blue solid lines) and the diagonal elements (red dashed lines) of the Hamiltonian $H_j$ due to the modulations shown in (a). In both figures, the vertical dotted lines indicate the positions at which the levels corresponding to the mode $j$ in both waveguides cross.}
    \label{fig:SCRAP-example}
\end{figure}

We study the implementation of the SCRAP method in a system of two planar evanescently coupled optical waveguides, each one supporting two TE modes. Modes of different order have their coupling suppressed both for energy and parity reasons, so one can describe the propagation of modes of the same order $j$ independently of the rest. The equation describing light propagation in such a system of two multimode waveguides then is:
\begin{equation}
    i\frac{d}{dz}\bm{a}_j(z) = H_j(z) \bm{a}_j(z), \label{SCRAP-analogSchrod}
\end{equation}
where $\bm{a}_j(z) = \left(a_j^{L}(z),a_j^{R}(z)\right)$ are the probability amplitudes of mode $j$ in the left and right waveguide, respectively. The Hamiltonian is defined as:
\begin{equation}
    H_j(z) = \begin{pmatrix}
        0 & c_j(z) \\
        c_j(z) & \Delta_j(z)
    \end{pmatrix},
    \label{SCRAP-Hz}
\end{equation}
where $c_j$ is the coupling between the modes $j$ of each waveguide and the detuning is the difference between the propagation constants in both waveguides, $\Delta_j(z) = \beta_j^{R}(z)-\beta_j^{L}(z) = \Delta_{j,0} - \Delta\beta_j(z)$, with $\Delta_{j,0}$ being the initial detuning and $\Delta\beta_j$ its variation along $z$. The coupling strength and the detuning play here the role of the Rabi frequency and the Stark pulse, respectively, in the original SCRAP implementation \cite{Yatsenko1999}. We will consider a Gaussian dependence on $z$ for both $c_j(z)$ and $\Delta\beta_j(z)$, as displayed in Fig.~\ref{fig:SCRAP-pulses}. The modulation of $\Delta\beta_j$ causes two level crossings at the points indicated by vertical dotted lines, see Fig.~\ref{fig:SCRAP-energy}, and we choose $c_j(z)$ to be strong during the first crossing and weak during the second.

One can diagonalize the Hamiltonian corresponding to each pair of modes $j$, Eq.~(\ref{SCRAP-Hz}), obtaining the following eigenvectors and eigenvalues:
\begin{gather}
    \psi_{j}^{-}(z) = \cos{\theta(z)} \phi_{j}^{L} - \sin{\theta(z)} \phi_{j}^{R}, \label{SCRAP-psi-} \\
    \psi_{j}^{+}(z) = \sin{\theta(z)} \phi_{j}^{L} + \cos{\theta(z)} \phi_{j}^{R}, \label{SCRAP-psi+} \\
    \lambda_{j}^{\pm} = \frac{\Delta_j(z)}{2} \pm \frac{1}{2} \sqrt{4c_j^2(z) + \Delta_j^2(z)}, \label{SCRAP-eigenE2}
\end{gather}
where $\phi_{j}^{L}$ and $\phi_{j}^{R}$ are the diabatic states in each waveguide, and $\tan{2\theta} = 2c_j(z)/\Delta_j(z)$ defines the mixing angle. Under the modulation of $c_j(z)$ and $\Delta\beta_j(z)$ in Fig.~\ref{fig:SCRAP-pulses}, the mixing angle evolves from $-\pi/2$ to \num{0}. Thus, if light is injected in the left waveguide, the system adiabatically follows $\psi_{j}^{+}$ from -$\phi_{j}^{L}$ to $\phi_{j}^{R}$, achieving complete light intensity transfer to the right waveguide. In Fig.~\ref{fig:SCRAP-energy} we show the typical variation along $z$ of the eigenvalues and the diagonal elements, i.e. the detuning $\Delta_j$, of the Hamiltonian in (\ref{SCRAP-Hz}). From that figure we see that an excitation in either mode follows the adiabatic level, being transferred into the other waveguide during the first level crossing where the coupling is maximal, and remaining unaltered during the second crossing where the coupling is negligible.

\subsection{Supersymmetry}

\begin{figure}[h]
    \centering
    \begin{subfigure}[t]{1\columnwidth}
    \includegraphics[width=\textwidth]{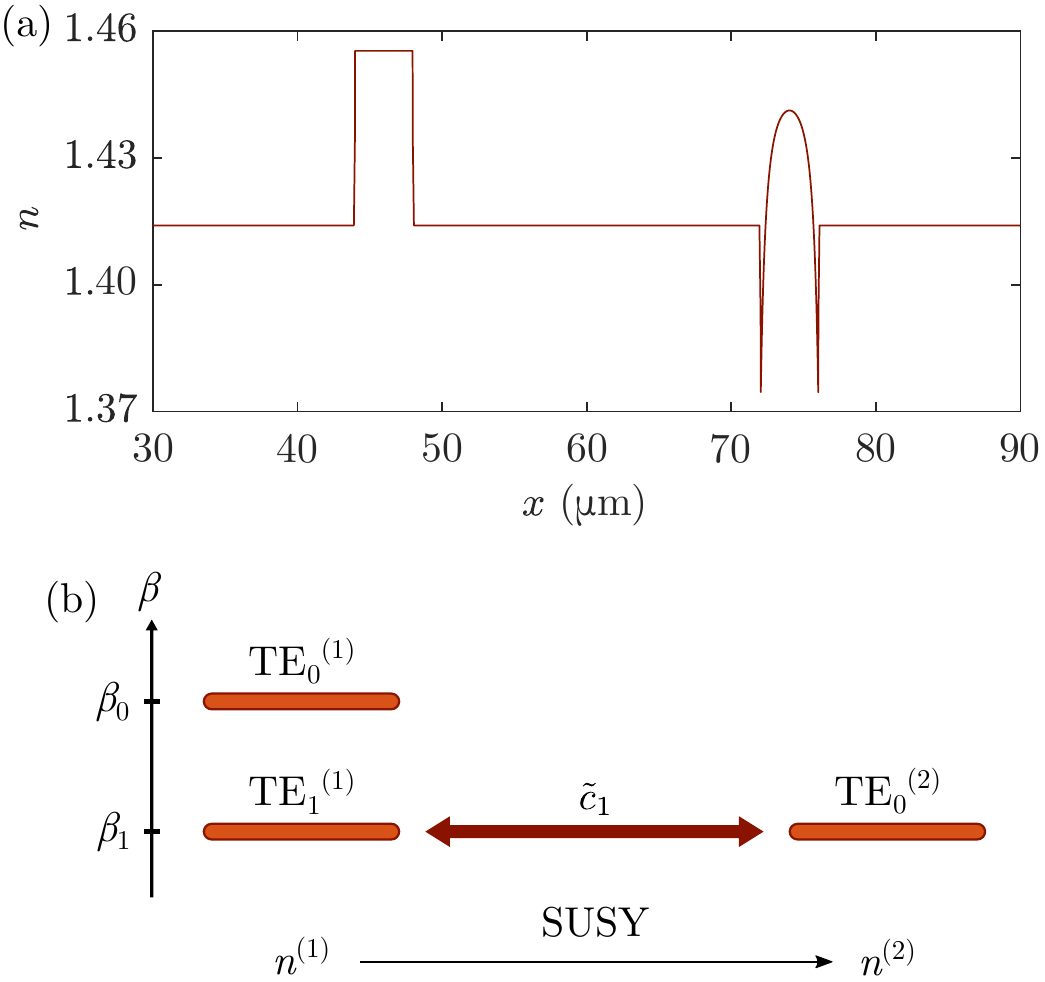}
    \phantomcaption
    \label{fig:SCRAP-SUSY-refindex-input}
    \end{subfigure}
    \begin{subfigure}[t]{0\columnwidth}
    \includegraphics[width=0\textwidth]{example-image-b}
    \phantomcaption
    \label{fig:SCRAP-SUSY-diagram}   
    \end{subfigure}
    \caption{\textbf{(a)} Refractive index profile of a step-index waveguide (left) and its superpartner profile (right). \textbf{(b)} Diagram of the propagation constants in a two-mode step-index waveguide defined by $n^{(1)}$ and in its superpartner profile $n^{(2)}$, where $\beta_0$ is eliminated from the spectrum. If both waveguides are brought close together, the fundamental mode from $n^{(2)}$, TE$_0^{(2)}$, only couples to the TE$_1^{(1)}$ mode of $n^{(1)}$ with strength $\tilde{c}_1$.}
    \label{fig:SCRAP-SUSY}
\end{figure}

In the context of optical waveguides, SUSY transformations allow one to construct a superpartner refractive index profile with the same modal content as the initial waveguide aside from the fundamental mode, which is not supported by the SUSY waveguide if the symmetry is unbroken. This allows for a great control of light transmission in multimodal waveguide structures, as only light in supported modes will be transferred to and from these waveguides. We show in Fig.~\ref{fig:SCRAP-SUSY-refindex-input} a step-index waveguide alongside its SUSY partner, and a diagram of its energy levels in Fig.~\ref{fig:SCRAP-SUSY-diagram}. The superpartner of a two-mode waveguide displays only a single guided mode, with a propagation constant matching the excited mode of the original waveguide. If we couple these two waveguides between them, only the modes with the same propagation constant will feel the coupling. That is, only the TE$_1^{(1)}$ and the TE$_0^{(2)}$ modes will be coupled.

To build a superpartner index profile one needs to compute the so-called superpotential, $W(x)$, which relates two superpartner Hamiltonians in the following way \cite{Cooper1995}:
\begin{align}
    \mathcal{H}^{(1)} &= -\frac{d^2}{dx^2} - \frac{dW}{dx} + W^2(x), \label{SCRAP-H1} \\
    \mathcal{H}^{(2)} &= -\frac{d^2}{dx^2} + \frac{dW}{dx} + W^2(x), \label{SCRAP-H2}
\end{align}
where in this case $\mathcal{H}^{(1)} = -\frac{d^2}{dx^2} - k_0^2 \left[n^{(1)}(x)\right]^2$ is the Hamiltonian from the optical Helmholtz equation \cite{saleh_teich_2007}. The superpotential can be computed from the electric field spatial distribution of the fundamental mode of the waveguide:
\begin{equation}
    W(x) = -\frac{d}{dx}\left[\ln{e_0^{(1)} (x)}\right], \label{SCRAP-superpot}
\end{equation}
which together with the definitions in (\ref{SCRAP-H1}) and (\ref{SCRAP-H2}) allows to compute the superpartner refractive index profile:
\begin{equation}
    n^{(2)}(x) = \sqrt{\left[n^{(1)}\right]^2-\frac{2}{k_0^2}\frac{dW}{dx}}. \label{SCRAP-nSUSY-general}
\end{equation}
Analytical expressions for $n^{(2)}(x)$ can only be derived for those cases in which exact solutions for the fundamental mode $e_0^{(1)}(x)$ can be found, otherwise the profile needs to be obtained numerically. For a step-index waveguide with core index $n_{core}$, the superpartner is defined by:
\begin{equation}
    n_{step}^{(2)} (x) = \sqrt{n_{core}^2-2\left(\frac{k_x}{k_0}\right)^2 \sec^2\left(k_x x\right)}, \label{SCRAP-SUSY-n}
\end{equation}
where $k_0$ is the vacuum wavenumber and $k_x = \sqrt{n_{core}^2k_0^2-\beta_0^2}$, with $\beta_0$ being the propagation constant of the fundamental mode of the step-index waveguide. This refractive index profile is the one shown in Fig.~\ref{fig:SCRAP-SUSY-refindex-input} alongside the original step-index profile.

\section{Physical implementation} \label{s-implementation}

\begin{figure}[h]
    \centering
    \includegraphics[width=\linewidth]{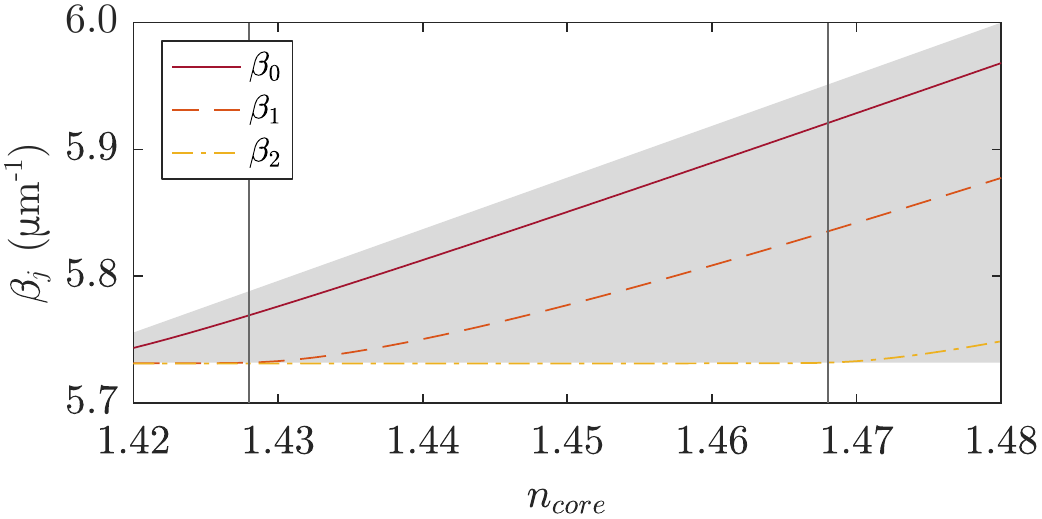}
    \caption{Variation of the propagation constants of the TE-modes of a step-index planar waveguide with respect to the core refractive index, for a width of $w = \SI{4}{\micro\meter}$, cladding index $n_{clad} = 1.414$ and wavelength $\lambda = \SI{1.55}{\micro\meter}$. The guided modes must have propagation constant values in the shaded region, $k_0 n_{clad} < \beta_j < k_0 n_{core}$. The waveguide is bi-modal in the region between vertical solid lines.}
    \label{fig:beta-n}
\end{figure}

Successful implementation of SCRAP requires the simultaneous control of the detuning and the coupling strength along the propagation direction. For laser-written waveguides \cite{Szameit2007}, the most precise approach to control their propagation constants experimentally is to modulate the core refractive index by adjusting the relative velocity between the sample and the laser. For a cladding refractive index of $n_{clad} = 1.414$, a width of $w = \SI{4}{\micro\meter}$ and a wavelength at the telecom range $\lambda = \SI{1.55}{\micro\meter}$, we show the variations of the propagation constants when changing the core index in Fig.~\ref{fig:beta-n}.
We see that the waveguides stop being bi-modal below $n_{core} \simeq 1.428$ and above $n_{core} \simeq 1.468$ as indicated by the vertical solid lines in Fig.~\ref{fig:beta-n}. From the same figure, however, we also observe that for values around $n_{core} \simeq 1.455$ the propagation constants for both the TE$_0^{(1)}$ and TE$_1^{(1)}$ modes are nearly linear in $n_{core}$, meaning that we can perform a linear fit for both in that region, while also maintaining only two supported modes in the waveguide. We obtain, for each mode $j$:
\begin{equation}
    \beta_j(n) = b_j n + q_j, \label{SCRAP-beta-n}
\end{equation}
with $b_0 = \SI{3.876}{\per\micro\meter}$, $q_0 = \SI{0.230}{\per\micro\meter}$, $b_1 = \SI{3.142}{\per\micro\meter}$ and $q_1 = \SI{1.222}{\per\micro\meter}$. Each propagation constant has a different variation with respect to the refractive index, but from now on we will only focus on the TE$_1^{(1)}$ mode.

\begin{figure}[h]
    \centering
    \includegraphics[width=\linewidth]{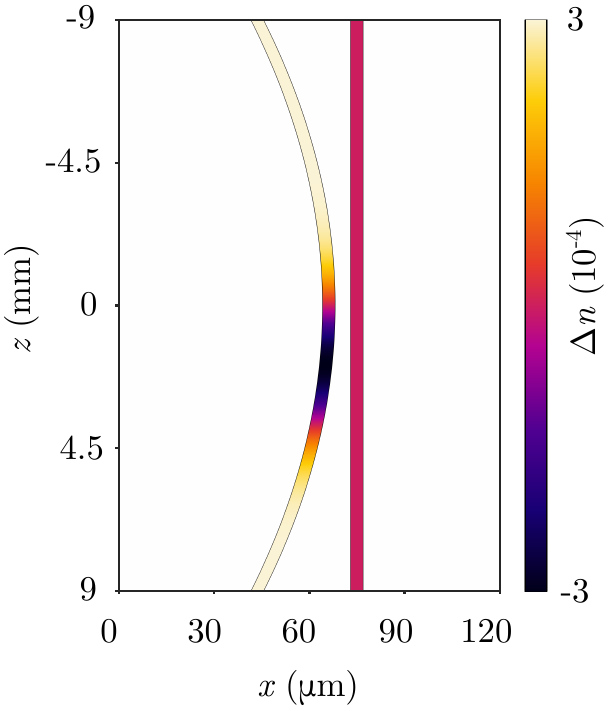}
    \caption{Geometry of the SCRAP implementation and refractive index of the waveguide cores. The modulation of the core of the left waveguide is represented as $\Delta n = n_{L} - n_{R}$.}
    \label{fig:SCRAP-refindex}
\end{figure}

We now consider the geometry showcased in Fig.~\ref{fig:SCRAP-refindex}, where we also display the refractive index modulation. The right waveguide features a constant core refractive index $n_R = 1.455$, while the refractive index of the left core is modulated following:
\begin{equation}
    n_L(z) = n_{L0} - 2(n_{L0}-n_R)\exp{\left(-\frac{(z-\zeta)^2}{Z_s^2}\right)}, \label{SCRAP-nz}
\end{equation}
where $n_{L0} = 1.4553$. Although the same effect can be achieved by performing the modulation on the right waveguide, which would actually correspond more naturally with the Hamiltonian in (\ref{SCRAP-Hz}), we choose to vary the left core for a direct comparison with the SUSY implementation described later. This Gaussian variation of the core index causes in turn a Gaussian change in the propagation constants through (\ref{SCRAP-beta-n}), and it is within current experimental reach \cite{Szameit2010}. $\zeta$ and $Z_s^2$ are parameters that determine the delay with respect to the coupling and the width of the Gaussian, respectively. We fix $Z_s = L/8$ and choose $\zeta = Z_s \sqrt{\log{2}}$ so that the first level crossing occurs at $z=0$, where the waveguides are closest and the coupling is maximal.

\begin{figure}[h]
    \centering
    \includegraphics[width=\columnwidth]{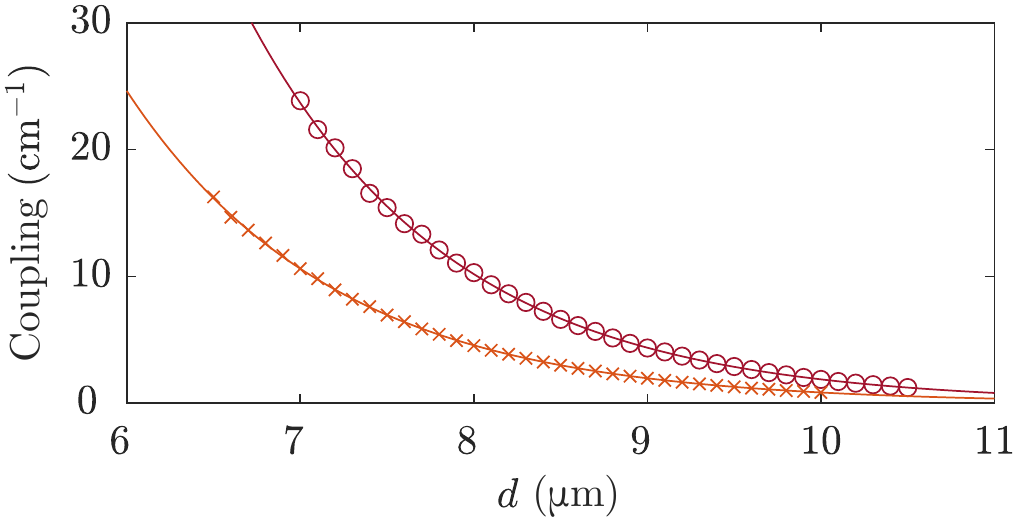}
    \caption{Computed coupling strength between the TE$_1^{(1)}$ modes of the two step-index waveguides (red circles), $c_1$, and between the TE$_1^{(1)}$ mode of the step-index waveguide and the TE$_0^{(2)}$ mode of the SUSY waveguide (orange crosses), $\tilde{c}_1$, with respect to the separation between each pair of waveguides. The exponential fit for each set of values is included in solid lines.}
    \label{fig:coupling}
\end{figure}

Both the coupling strength and the transfer efficiency are reduced as $n_L$ deviates from $n_R$ \cite{saleh_teich_2007}. However, the relative variation for the coupling between the central value $n_L = 1.455$ and the limits of the modulation has been computed to be below \num{e-7} for the parameters considered in this work. As such, we can consider that the coupling does not deviate from the case of equal waveguides. In that case, the coupling strength between modes can simply be controlled by adjusting the distance between the waveguides, since the coupling decreases exponentially with their separation: 
\begin{equation}
    c_j(d) = c_{j}^{0}\exp\left(-\kappa_j d\right). \label{SCRAP-cd}
\end{equation}
Knowing this, we perform a numerical study to determine the parameters $c_{j}^{0}$ and $\kappa_j$ for the TE$_1$ modes of identical waveguides with $n_{core} = 1.455$. The coupling strength for each distance can be gathered by means of the beat length:
\begin{equation}
    L_j = \frac{\pi}{2c_j},
\label{SCRAP-beatlength}    
\end{equation}
which corresponds to the length at which complete light transfer occurs. By performing simulations at different distances and measuring the beat length we obtain the distribution in Fig.~\ref{fig:coupling}, from which we extract the following parameters: $c_{1}^{0} = \SI{8707.8}{cm^{-1}}$ and $\kappa_1 = \SI{0.844}{\per\micro\meter}$.

For the geometry in Fig.~\ref{fig:SCRAP-refindex}, the coupling along $z$ approximately follows a Gaussian expression \cite{Queralto2017}: 
\begin{equation}
    c_j(z) \approx c_j(0) \exp{\left(-\kappa_j\frac{z^2}{2r}\right)}, \label{SCRAP-cz}
\end{equation}
whose width can be controlled through the curvature radius $r$. In there, $c_j(0)$ corresponds to the coupling at the point where the waveguides are closest, as computed from (\ref{SCRAP-cd}) for $d = x_{m}$.

To describe the implementation of SUSY, let us come back to the situation depicted in Fig.~\ref{fig:SCRAP-SUSY}. The propagation constants of the fundamental mode of the SUSY waveguide, TE$_0^{(2)}$, and the TE$_1^{(1)}$ mode of the step-index waveguide will be equal, but the former will be much less extended into the cladding region than the latter. As such, compared to the coupling strength between the TE$_1^{(1)}$ modes of two step-index waveguides, the coupling between SUSY and step-index waveguides will be weaker. To prove this, we measure the beat length between the TE$_1^{(1)}$ and TE$_0^{(2)}$ modes at different distances, and compute their coupling strength from (\ref{SCRAP-beatlength}). We plot the results in Fig.~\ref{fig:coupling} and perform an exponential fit following Eq.~(\ref{SCRAP-cd}), from which we can extract $\tilde{c}_{1}^0 = \SI{3855.8}{cm^{-1}}$ and $\tilde{\kappa}_1 = \SI{0.842}{\per\micro\meter}$. We denote the parameters with a tilde to indicate that they correspond to the coupling between the excited mode of the step-index waveguide and the fundamental mode of its superpartner profile.

To add SUSY to the SCRAP implementation, the most convenient method is to apply the modulation and the SUSY transformation to different waveguides. Hence, we substitute the right waveguide by its superpartner index profile, computed using the expression in (\ref{SCRAP-SUSY-n}) with a constant $n_{core} = 1.455$. This results in a refractive index profile at the input facet similar to the one shown in Fig.~\ref{fig:SCRAP-SUSY-refindex-input}, but with the left core having an initial refractive index of $n_{L0} = 1.4553$. This core is then modulated along $z$ in the same way as in Fig.~\ref{fig:SCRAP-refindex}.

\section{Results} \label{s-results}

We apply SCRAP combined with SUSY to efficiently transfer light intensity between multimode waveguides. First, in section~\ref{pumping} we exploit this efficient transfer to pump the excited mode of the step-index waveguide. The SUSY waveguide is single-mode, and as such its guided mode is easy to excite. This excitation can only be transferred to the TE$_1^{(1)}$ mode due to the phase matching condition, thus achieving efficient pumping of this mode. To quantify the accuracy of the method, we define a fidelity:
\begin{equation}
    F=\left|\left<\phi_t|\psi_{out}\right>\right|^2, \label{SCRAP-stateFidelity}
\end{equation}
which is an integral that compares the spatial profile of the output mode $\psi_{out}$ with the one of a particular target mode $\phi_t$.

After that, we consider the demultiplexing possibilities of the method. We first consider the SCRAP technique without applying SUSY in section~\ref{demult-nosusy}, and then comment on the improvement that SUSY brings in section~\ref{demult-susy}. Since the light intensity of the TE$_1^{(1)}$ mode is efficiently transferred between waveguides but the one for the TE$_0^{(1)}$ is not, SCRAP allows to spatially separate a superposition of both. For this other application, we define the following figure of merit:
\begin{equation}
    \mathcal{F} = \frac{I_{0,\text{out}}^{L}}{I_{0,\text{in}}^{L}}\cdot\frac{I_{1,\text{out}}^{R}}{I_{1,\text{in}}^{L}}, \label{SCRAP-FofMerit}
\end{equation}
which is just the fraction of intensity of the TE$_0^{(1)}$ mode that remains on the input waveguide multiplied by the fraction of intensity of the TE$_1^{(1)}$ mode that gets transferred either to the other TE$_1^{(1)}$ mode in section~\ref{demult-nosusy} or to the TE$_0^{(2)}$ mode in section~\ref{demult-susy}.

\subsection{Pumping of the TE$_1^{(1)}$ mode} \label{pumping}

\begin{figure}[h]
    \centering
    \includegraphics[width=\linewidth]{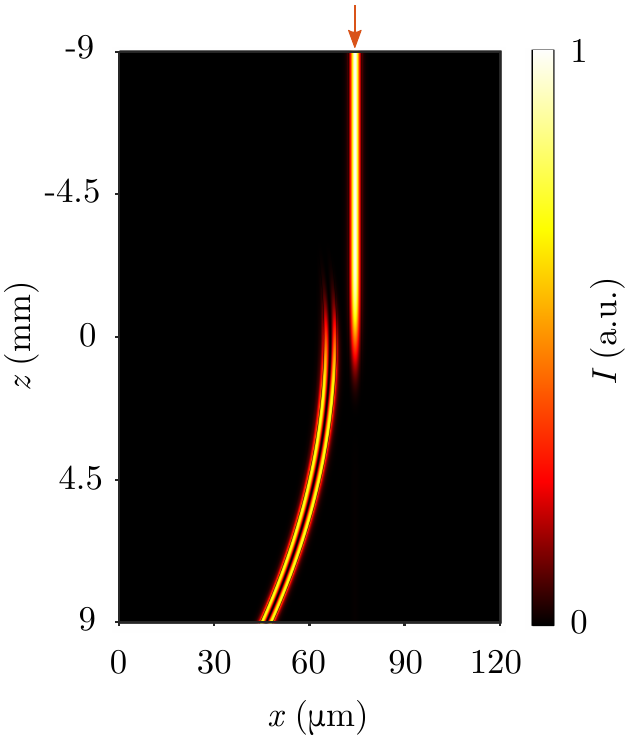}
    \caption{Light intensity propagation for the pumping of the TE$_1^{(1)}$ mode of the step-index waveguide in the SUSY implementation of SCRAP.}
    \label{fig:SCRAP-SUSY-pumpTE1}
\end{figure}

We perform finite difference numerical simulations to demonstrate that it is possible to efficiently pump the TE$_1^{(1)}$ mode of the step-index waveguide starting from an excitation of the fundamental mode TE$_0^{(2)}$ of the SUSY waveguide, which is now single-mode as depicted in Fig.~\ref{fig:SCRAP-SUSY-diagram}. We show the light intensity propagation of this process in Fig.~\ref{fig:SCRAP-SUSY-pumpTE1}.

\begin{figure}[h]
    \centering
    \includegraphics[width=\columnwidth]{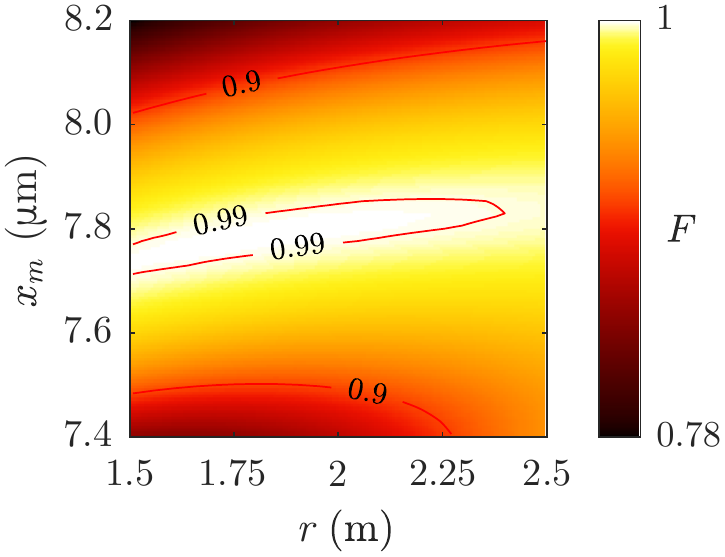}
    \caption{Fidelity of the pumping of the TE$_1^{(1)}$ mode of the step-index waveguide when propagating an excitation on the fundamental mode of the SUSY waveguide, as a function of the minimum distance between waveguides, $x_{m}$, and the radius of curvature of the left waveguide, $r$.}
    \label{fig:SCRAP-F(TE1)}
\end{figure}

We compute the fidelity of the pumping according to (\ref{SCRAP-stateFidelity}), and show in Fig.~\ref{fig:SCRAP-F(TE1)} that we can obtain the target mode with fidelities above $F=0.9$ for a wide range of parameter values, with a significant region where fidelities exceed $F=0.99$. Even if the geometrical parameters are slightly off and power transfer is not complete, light on the step-index waveguide will almost entirely be comprised by the TE$_1^{(1)}$ mode. This results from the fact that despite their different spatial profiles, the SUSY fundamental mode can only be efficiently coupled to the TE$_1^{(1)}$ mode of the step-index waveguide due to the phase-matching condition, and thus this procedure produces a negligible excitation of the TE$_0^{(1)}$ mode. This can be proved by computing the overlap between the output field and the TE$_0^{(1)}$ mode of the step-index waveguide, in the same way as in (\ref{SCRAP-stateFidelity}), which yields values below $\num{1.3e-3}$ for all parameters considered in Fig.~\ref{fig:SCRAP-F(TE1)}.

\begin{figure}[h]
    \centering
    \includegraphics[width=\linewidth]{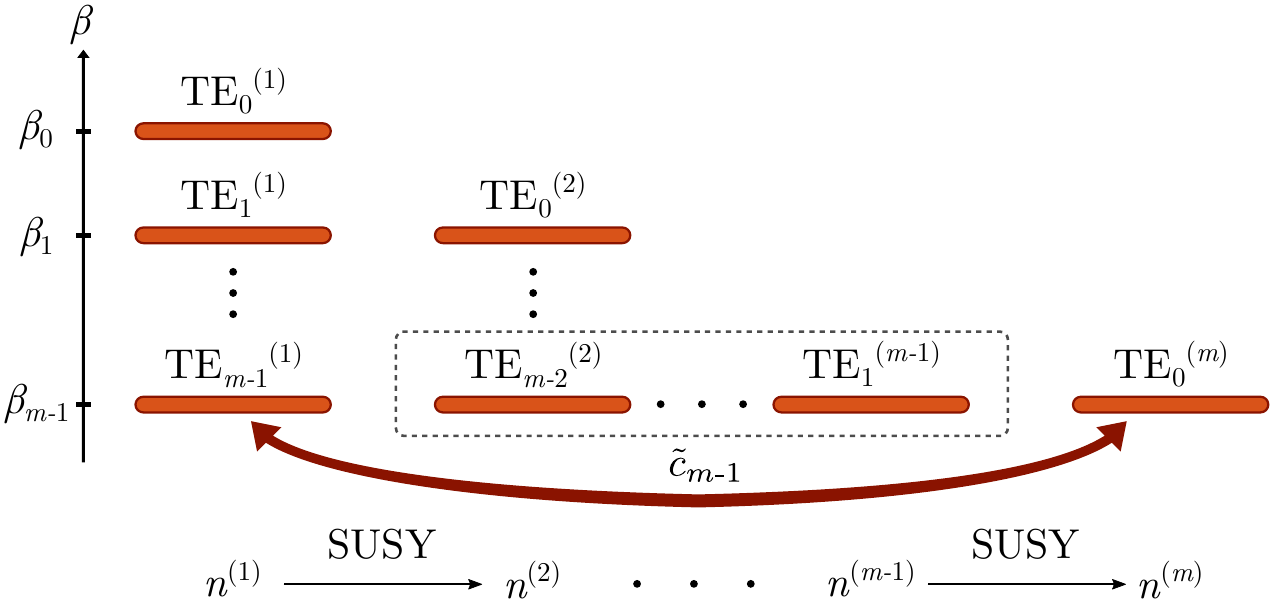}
    \caption{Diagram of propagation constants in a $m$-mode step-index waveguide (left column) and its superpartners after successive SUSY transformations. After $m-1$ transformations, only a single guided mode remains. Coupling this $n^{(m)}$ superpartner with the original structure should only yield significant coupling between this fundamental mode and the $(m-1)$-th mode of the original structure, with strength $\tilde{c}_{m-1}$.}
    \label{fig:SCRAP-SUSY-diagram-m}
\end{figure}

This scheme can also be applied to pump higher-order modes in higher-order multimode waveguides. We show a diagram of the effect of applying successive SUSY transformations to a $m$-mode step-index waveguide in Fig.~\ref{fig:SCRAP-SUSY-diagram-m}. One can couple the $m$-mode waveguide to its superpartner after $m-1$ consecutive SUSY transformations, which is single mode. Modulating the core of the $m$-mode waveguide allows to implement the SCRAP scheme between its TE$_{m-1}^{(1)}$ mode, and the single mode of the SUSY waveguide. With this, one can pump the highest order mode of the original waveguide using a fundamental-mode input on its SUSY partner.

\begin{figure*}[t]
    \centering
    \begin{subfigure}[t]{1\textwidth}
    \includegraphics[width=\textwidth]{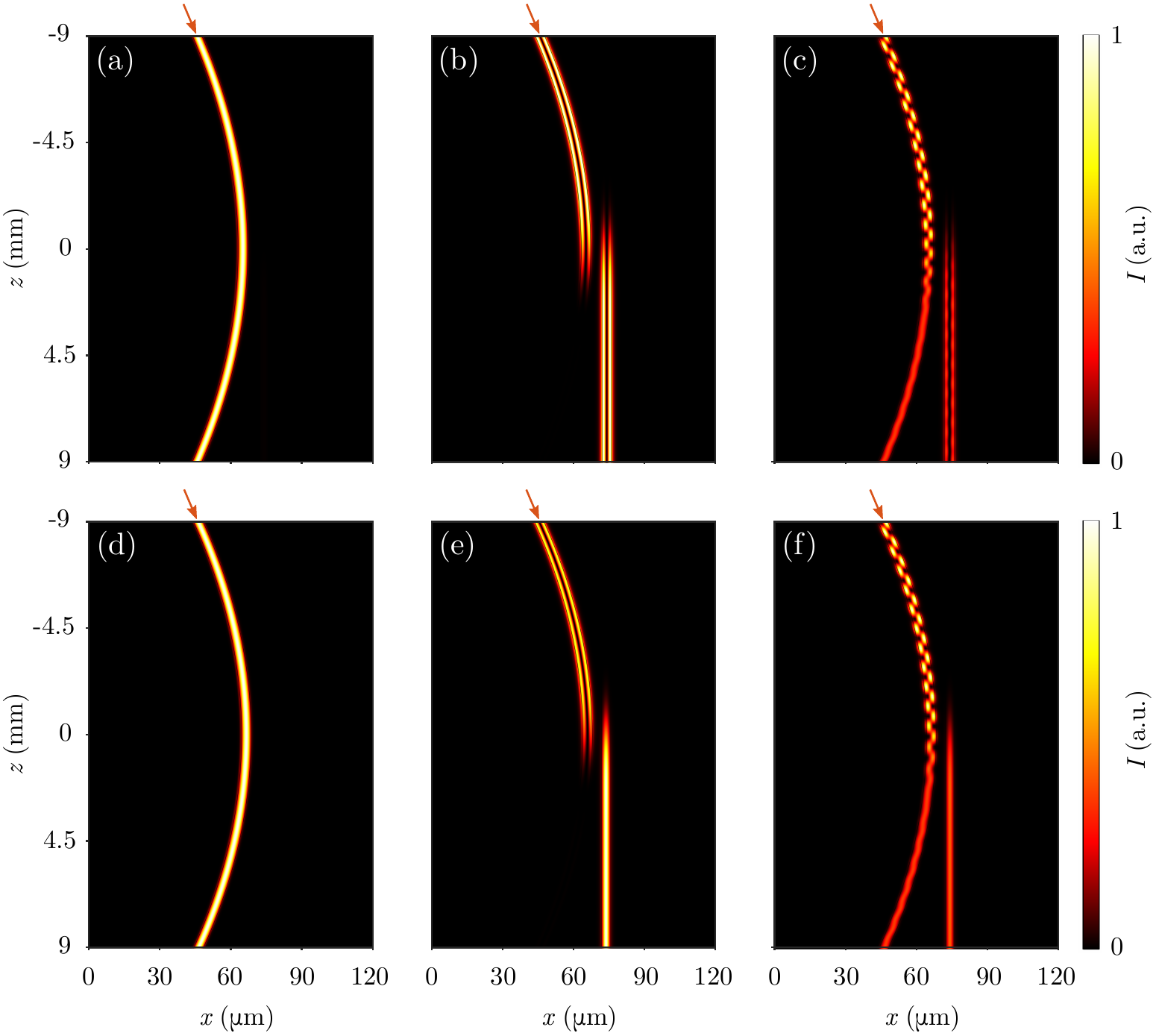}
    \phantomcaption
    \label{fig:SCRAP-TE0}
    \end{subfigure}
    \begin{subfigure}[t]{0\columnwidth}
    \includegraphics[width=0\textwidth]{example-image-b}
    \phantomcaption
    \label{fig:SCRAP-TE1}   
    \end{subfigure}
    \begin{subfigure}[t]{0\columnwidth}
    \includegraphics[width=0\textwidth]{example-image-b}
    \phantomcaption
    \label{fig:SCRAP-both}   
    \end{subfigure}
    \begin{subfigure}[t]{0\textwidth}
    \includegraphics[width=0\textwidth]{example-image-b}
    \phantomcaption
    \label{fig:SCRAP-SUSY-TE0}
    \end{subfigure}
    \begin{subfigure}[t]{0\columnwidth}
    \includegraphics[width=0\textwidth]{example-image-b}
    \phantomcaption
    \label{fig:SCRAP-SUSY-TE1}   
    \end{subfigure}
    \begin{subfigure}[t]{0\columnwidth}
    \includegraphics[width=0\textwidth]{example-image-b}
    \phantomcaption
    \label{fig:SCRAP-SUSY-both}   
    \end{subfigure}
    \caption{Comparison between the light intensity propagation in the SCRAP implementation between to step-index waveguides (top row) and the SUSY-enhanced implementation (bottom row) when injecting \textbf{(a), (d)} the TE$_0^{(1)}$ mode \textbf{(b), (e)} the TE$_1^{(1)}$ mode and \textbf{(c), (f)} an equally-weighted superposition of both in the left waveguide.}
    \label{fig:regular-SCRAP}
\end{figure*}

\subsection{Mode demultiplexing} \label{demult-nosusy}

As previously mentioned, the SCRAP scheme allows to efficiently transfer light intensity to the TE$_1^{(1)}$ mode between step-index waveguides, which can be exploited for mode demultiplexing even without applying SUSY. We first consider the geometry in Fig.~\ref{fig:SCRAP-refindex} and perform numerical simulations for the propagation of the TE$_0^{(1)}$ mode, the TE$_1^{(1)}$ mode and an equally-weighted superposition of both. We display the results in Figs.~\ref{fig:SCRAP-TE0}, \ref{fig:SCRAP-TE1} and \ref{fig:SCRAP-both}, respectively. From those figures, we can readily see that light in the TE$_1^{(1)}$ mode is efficiently transferred, while the TE$_0^{(1)}$ mode remains mostly confined in the left waveguide. The variation of the refractive index causes different mode levels to vary differently, see Fig.~\ref{fig:beta-n}. Also, since the spatial profile of the TE$_0^{(1)}$ mode is less extended into the cladding region compared to more excited states \cite{saleh_teich_2007}, the coupling between fundamental modes is weaker than that for the TE$_1^{(1)}$ modes. Both these reasons cause the transmission for the TE$_0^{(1)}$ mode to be less efficient when compared to the TE$_1^{(1)}$ mode. Nonetheless, a small fraction of its power is still transferred, and that hinders the efficiency of the device as a demultiplexer. This can be seen most easily in Fig.~\ref{fig:SCRAP-both}, where the beating in intensity implies that the right waveguide holds a superposition of both TE modes, with a relative fraction that depends on the strength of the coupling for the TE$_0^{(1)}$ mode.

\begin{figure}[h]
    \centering
    \begin{subfigure}[t]{1\columnwidth}
    \includegraphics[width=\textwidth]{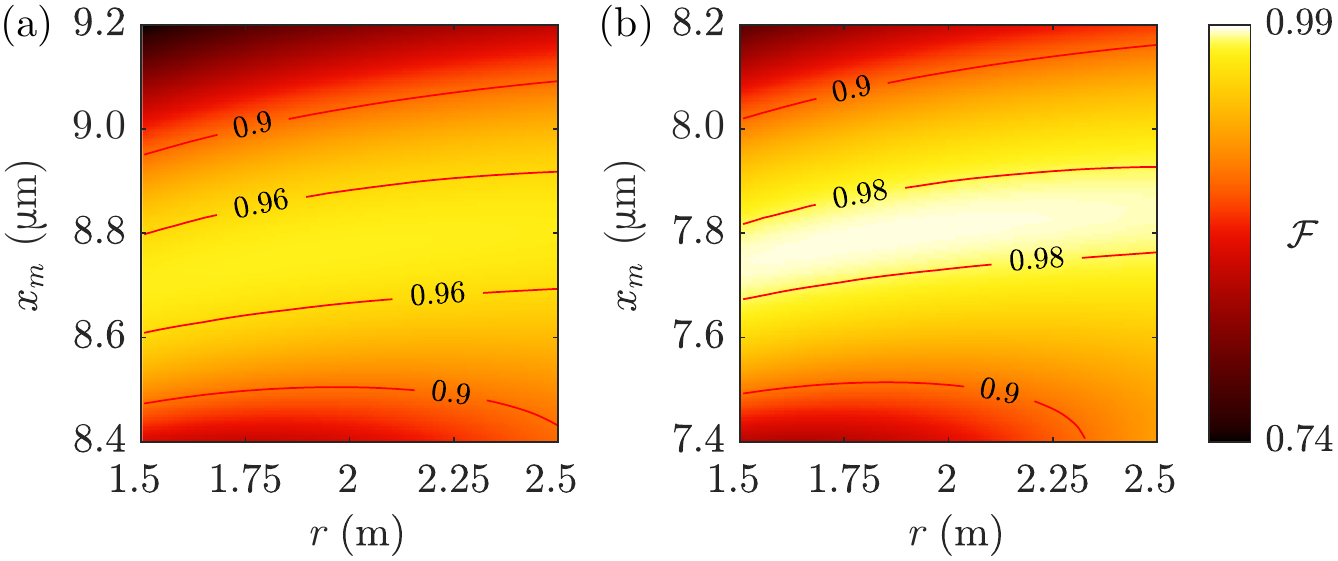}
    \phantomcaption
    \label{fig:SCRAP-fid-demult-normal}
    \end{subfigure}
    \begin{subfigure}[t]{0\columnwidth}
    \includegraphics[width=0\textwidth]{example-image-b}
    \phantomcaption
    \label{fig:SCRAP-fid-demult-susy}   
    \end{subfigure}
    \caption{Figure of merit defined in (\ref{SCRAP-FofMerit}) for the spatial separation of the TE$_0^{(1)}$ and TE$_1^{(1)}$ modes in \textbf{(a)} the step-index SCRAP implementation and \textbf{(b)} the SUSY implementation as a function of the minimum distance between waveguides, $x_m$, and the radius of curvature, $r$. We see that the efficiencies are overall higher in the SUSY case.}
    \label{fig:SCRAP-fid-demult}
\end{figure}

To check exactly how efficient the method is, we compute the figure of merit defined in (\ref{SCRAP-FofMerit}), and plot the results in Fig.~\ref{fig:SCRAP-fid-demult-normal}. We see that there is a wide region where $\mathcal{F} > 0.9$, but it caps at around $\mathcal{F} = 0.97$.

\subsection{SUSY-enhanced demultiplexing}  \label{demult-susy}

Replacing the right waveguide with its SUSY partner entirely removes the possibility of coupling for the TE$_0^{(1)}$ mode, since its equivalent on the SUSY waveguide is no longer supported, whilst maintaining the possibility of implementing the SCRAP efficiently for the TE$_1^{(1)}$ mode. 

We repeat the finite difference numerical simulations, and show the results of inputting the TE$_0^{(1)}$, TE$_1^{(1)}$ and a superposition of both modes in Figs.~\ref{fig:SCRAP-SUSY-TE0}, \ref{fig:SCRAP-SUSY-TE1} and \ref{fig:SCRAP-SUSY-both}, respectively. In this case, light in the TE$_0^{(1)}$ mode remains entirely confined in the left waveguide, while power is efficiently transferred between the TE$_1^{(1)}$ and the TE$_0^{(2)}$ of the SUSY partner waveguide. This has favorable implications in the efficiency, as shown in Fig.~\ref{fig:SCRAP-fid-demult-susy}. The figure of merit given in Eq.~(\ref{SCRAP-FofMerit}) is overall larger for all parameter values, and reaches a maximum of $\mathcal{F} = 0.99$ for $x_m = \SI{7.8}{\micro\meter}$ and a radius of curvature around $r = \SI{2}{\meter}$.

However, the intensity profile of the TE$_1^{(1)}$ mode is transformed into that of the fundamental mode of the right waveguide upon transfer, and is no longer antisymmetric. Introducing a superposition of modes into the device indeed leads to efficient mode separation but alters the intensity profile of the excited mode, causing the output to resemble the one from a beam splitter instead of the one from a traditional demultiplexer. In principle, adding a third bent waveguide on the right side should allow one to perform traditional demultiplexing \cite{Queralto2017} by implementing the three-state extension of the SCRAP method discussed in \cite{Rangelov2005}. In that case, the spatial profile of the TE$_1^{(1)}$ waveguide is recovered and the suppression of the transfer for the TE$_0^{(1)}$ mode is maintained.

\section{Conclusions} \label{s-conclusions}

In this work, we have described how we can implement SCRAP in a system of multimode waveguides by modulating the core refractive index of one of the waveguides, and further enhance its performance by applying a SUSY transformation to the refractive index profile of the other one. An excitation on the fundamental mode of the SUSY waveguide, single-mode, gets transferred and pumps the excited mode of the modulated step-index waveguide with very high efficiency, reaching fidelities above $F=0.99$ in a wide region of parameter values.

The SCRAP technique can also be exploited for mode demultiplexing in step-index multimode waveguides. The TE$_0^{(1)}$ mode has a weaker coupling to the other waveguide than the exited mode, and light intensity on this mode remains mostly confined on the input waveguide, while light on the TE$_1^{(1)}$ mode is efficiently transferred. One can exploit this to demultiplex a superposition of these modes, or to obtain an equally-weighted superposition from individual excitations on each mode by reversing the device. Computing the figure of merit for the demultiplexing process, we see that the method easily exceeds $\mathcal{F} = 0.9$, but struggles to reach very high values. The efficiency of this process can be enhanced by substituting one of the waveguides by its SUSY partner. Unlike the previous case, the TE$_0^{(1)}$ mode of the step-index waveguide is not supported by the SUSY one, implying that it cannot couple to it, while light in the TE$_1^{(1)}$ mode can still be efficiently transferred into the SUSY fundamental mode by virtue of the SCRAP scheme. This allows for mode spatial separation with the figure of merit reaching $\mathcal{F} = 0.99$, and overall with higher values than without SUSY, at the cost of altering the spatial profile of the transferred mode. Additionally, this method enables the possibility of turning fundamental-mode excitations on both waveguides into a superposition of fundamental and excited modes of the step-index waveguide.

The technological difficulty of implementing the SUSY refractive index profile can be lowered by considering a super-Gaussian profile for the initial waveguide, whose superpartners display much softer profiles than the ones for a step-index waveguide \cite{Miri2014Proc} while having a similar modal content. Higher-order superpartners allow to pump higher-order excited modes of the original waveguide, which are traditionally very challenging to excite precisely. The technique also goes beyond the context of optical waveguides. It would be of interest to extend the SCRAP implementation to different physical systems, as any system with multiple controllable energy levels can benefit from the technique.

\begin{acknowledgments}
The authors acknowledge financial support from the Spanish Ministry of Science and Innovation MICINN (contracts no. FIS2017-86530-P and PID2020-118153GB-I00) and Generalitat de Catalunya (contract no. SGR2017-1646). We also thank Gerard Queralt\'{o} for helpful discussions.
\end{acknowledgments}

\bibliographystyle{unsrt}
\bibliography{main}

\end{document}